\documentclass[a4paper, oneside, twocolumn, notitlepage, 10pt]{extarticle_ecoc}
\usepackage{ecoc}
\usepackage{myecoc}
\usepackage{subcaption}

\begin{document}
    \selectlanguage{english} 


    \title{Mitigating Outages in a 4.6-km FSO Link via Mode-Diverse Reception: An Experimentally Validated Digital Twin Approach}%
    

    \author{ Jonas Krimmer\textsuperscript{(1)}, Vincent van Vliet\textsuperscript{(2)}, Menno van den Hout\textsuperscript{(2)}, Eduward Tangdiongga\textsuperscript{(2)},\\Chigo Okonkwo\textsuperscript{(2)}, Sebastian Randel\textsuperscript{(1)} }
    \maketitle 


    \begin{strip}
        \begin{author_descr}

            \textsuperscript{(1)} Institute of Photonics and Quantum Electronics (IPQ), Karlsruhe Institute of Technology (KIT), Karlsruhe, Germany. \textcolor{blue}{\uline{jonas.krimmer@kit.edu}}%

            \textsuperscript{(2)} Electro-Optical Communication Group, Eindhoven University of Technology, the Netherlands
        \end{author_descr}%
    \end{strip}

    \renewcommand{\footnotemark}{}
    \renewcommand{\footnoterule}{}


    \begin{strip}
        \begin{ecoc_abstract}
            Terrestrial coherent FSO requires mitigating turbulence-induced coupling losses. We experimentally validate a \enquote{Digital Twin} channel model against fading statistics from a 4.6-km link. Combining long-term scintillometer data with mode-diverse receiver simulations, we demonstrate that a 6-mode receiver reduces turbulence-induced outage probabilities to \qty{2.02e-5}{}. © 2026 The Author(s)
        \end{ecoc_abstract}
    \end{strip}


    \section{Introduction}
    The evolution toward 6G mobile wireless networks and the massive bandwidth demands of artificial-intelligence-based applications require seamless optical backhaul across both fiber and wireless domains. Consequently, free-space optical (FSO) communication has the potential to become a critical building block of next-generation communication networks, offering fiber-like capacities and low latencies to bridge the gap between fixed fiber backhaul and dynamic terrestrial or non-terrestrial endpoints~\cite{prasadtera6GOverviewNext2025}.

    In spite of these advantages, atmospheric channel impairments severely limit the performance of terrestrial FSO links. While macroscopic weather phenomena such as fog and rain induce slowly evolving attenuation, allowing for mitigation using appropriate static link margins or automatic gain control, atmospheric turbulence induces rapid, spatio-temporal refractive index fluctuations~\cite{Andrews2005LaserEdition}. These millisecond-scale wavefront distortions potentially cause deep, catastrophic signal fades that exceed practical forward error correction (FEC) interleaving depths and force coherent digital signal processing (DSP) tracking loops to re-lock~\cite{valjusReviewAnalysisDigital2025}. Hence, these phase distortions severely impair coupling into a single-mode fiber (SMF), which is inevitable for integrating FSO links into fiber-based coherent optical networks.

    To mitigate this catastrophic coupling loss, mode-diverse reception schemes utilizing few-mode fibers (FMF) paired with multi-plane light conversion (MPLC) demultiplexers or photonic lanterns have been proposed~\cite{Fontaine2019}. However, modal diversity introduces a severe system trade-off: every additional supported spatial mode demands a dedicated coherent receiver and strongly scales the required DSP complexity. While theoretical works assessing the mode requirements exist~\cite{krimmerStatisticalAnalysisFreeSpacetoFiber2020}, system designers lack experimentally validated guidelines to determine how many modes are needed to guarantee specific long-term reliability targets under real-world atmospheric conditions.

    To address this validation gap, we propose a \enquote{Digital Twin} FSO channel model, rigorously validated against the deployed 4.6-km-long terrestrial Reid Photonloop link in Eindhoven, The Netherlands. First, we demonstrate that our numerical simulation aligns with the experimentally measured amplitude and phase fading statistics across distinct turbulence regimes. Subsequently, by integrating these validated fields with long-term scintillometer data, we quantify the engineering trade-off between mode-count complexity and long-term outage probability, providing a validated design framework for robust mode-diverse FSO receivers.

    \section{Experimental Setup and Digital Twin}
\begin{figure*}[t]
        \includegraphics[width=1\linewidth]{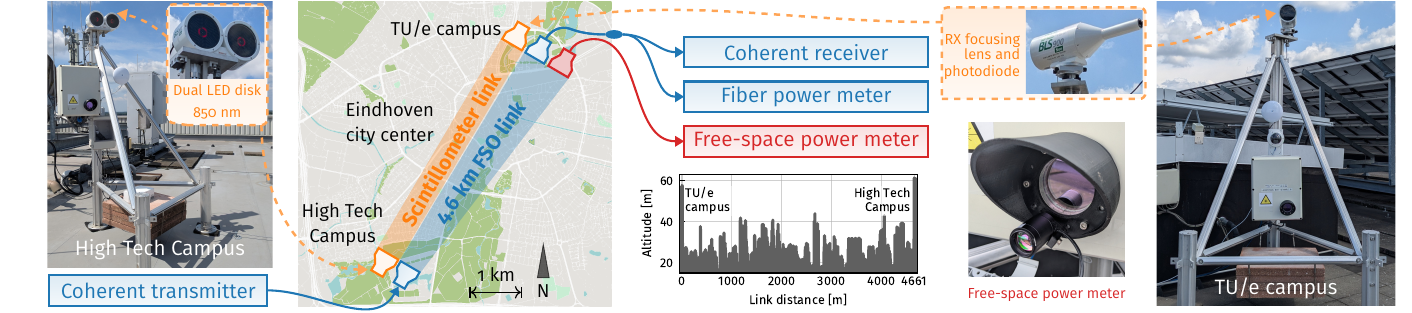}
        \caption{Experimental setup of the 4.6 km FSO link, illustrating the simultaneous large-aperture free-space measurement and SMF coupling branches alongside the scintillometer.}
        \label{fig:setup}
    \end{figure*}

    The experimental validation relies on the Reid Photonloop, a permanently established 4.6-km-long FSO testbed in Eindhoven, The Netherlands. As illustrated in \autoref{fig:setup}, the link connects the Eindhoven University of Technology campus to the High Tech Campus via fiber-coupled terminals (developed by Aircision) mounted on rooftop infrastructure. Traversing the city center at an altitude of roughly \qty{40}{\m}, the low elevation and highly irregular urban surface profile promote strong turbulence with significant contributions from both inertial and dissipation subranges, making it highly representative of critical terrestrial deployments such as cellular backhaul \cite{vanVliet_JLT}. Each terminal features automated pointing and tracking, continuously optimizing the free-space-to-fiber coupling at an operating wavelength of $\lambda = \qty{1560}{\nm}$.

    To provide atmospheric telemetry, we operate a scintillometer (Scintec BLS900 Neo) in parallel that calculates the path-averaged refractive index structure constant $C_{n}^{2}$ utilizing a 10-minute integration window. From the cumulative distribution function (CDF) of this long-term data, we isolate representative atmospheric states: the median (\qty{1.5e-15}{\m^{-2/3}}), 90\textsuperscript{th} percentile (\qty{7.3e-15}{\m^{-2/3}}), and the 99\textsuperscript{th} percentile (\qty{1.5e-14}{\m^{-2/3}}) turbulence strengths. This classification provides the reference points for the verification of our simulation model.

    At the receiver, a beam splitter routes the communication path into a single-mode fiber monitored by a high-speed optical power meter (Keysight N7745-C). Simultaneously, a spatially adjacent lens (Thorlabs LA1951-C) with diameter $D_{\text{FS,PM}}=\qty{25.4}{\mm}$ and focal length $f_{\text{FS,PM}}=\qty{25}{\mm}$ focuses the free-space field onto a \qty{9.7}{\mm} active-area detector (Thorlabs S122C). Even under 99th-percentile turbulence, the root-mean-square focal plane spot displacement induced by angle-of-arrival fluctuations
    is orders of magnitude smaller than the detector area~\cite{Andrews2005LaserEdition}. Consequently, this secondary branch captures nearly all of the light passing the 1-inch pupil, isolating aperture-averaged amplitude fluctuations to serve as an amplitude-only calibration target.

    To construct the digital twin, we simulate wave propagation through the turbulent channel using the split-step Fourier method with highly accurate phase screens adhering to the Frehlich spectrum~\cite{frehlichLaserScintillationMeasurements1992a,krimmerNewApproachNonuniform2020}. The propagation path $L$ is discretized into 30 steps of size $\Delta z$ to satisfy sampling requirements~\cite{schmidtNumericalSimulationOptical2010}. We iteratively compute the field's complex amplitude via $\ubar{u}_{k+1}= \mathcal{\ubar D}_{\Delta z}\mathcal{\ubar R}_{k}u_{k}$, where the operator $\mathcal{\ubar R}_{k}$ applies the turbulence-induced phase distortions, and the operator $\mathcal{\ubar D}_{\Delta z}$ models the (paraxial) propagation. To prevent aliasing at the receiver plane, we enforce a transverse grid resolution of \qty{1}{\mm} with $256 \times 256$ samples and apply super-Gaussian windowing~\cite{schmidtNumericalSimulationOptical2010}. Based on the link altitude, we assume an outer turbulence scale of \qty{10}{\m} and an inner scale of \qty{5}{\mm} \cite{mccraeInvestigatingOuterScale2019,ochsOpticalscintillationMethodMeasuring1985}. Applying the Monte Carlo method, i.\,e., repeating this split-step simulation for a large number of channel realizations (\(N_\text{iter}=\qty{10000}{}\)), we can evaluate statistical information from the received complex amplitude distribution at the receiver.

    Fiber coupling is simulated by evaluating the overlap integral between the received complex amplitude within the receiver aperture and the backpropagated fundamental SMF mode. The pointing-and-tracking feature of the Aircision terminals is considered by numerically removing the wavefront tilt. To validate the twin, both the simulated SMF coupling efficiency ($\eta$) and the simulated free-space intensity integrated over the 1-inch aperture are normalized to their respective means, $\eta_{\text{n}}= \eta / \langle\eta\rangle$.

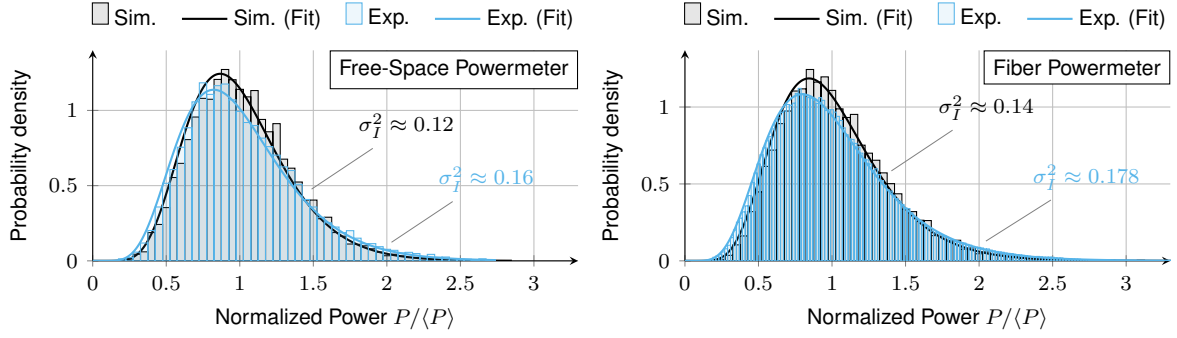
\begin{figure*}[t]
\begin{tikzpicture}
    \def\histogramfile{figures/histograms/01_Cn2_1.5e-15_waveprop_power_in_aperture.csv}
    \def\experimentfile{figures/histograms/2026-01-27T03-50-00_free_space_power_histogram.csv}
    \tikzset{font=\footnotesize}
    \begin{axis}[
        width=0.5*\linewidth, 
        height=0.6*\axisdefaultheight, 
        axis lines={left}, 
        xlabel={{Normalized Power $P / \langle P\rangle$}}, 
        ylabel={{Probability density}}, 
        enlarge x limits={{rel=0.1, upper}}, 
        enlarge y limits={{rel=0.1, upper}}, 
        xmin=0, 
        xmax=3,
        ymin={0}, 
        grid, 
        legend columns={-1}, 
        legend style={{ /tikz/every even column/.append style={column sep=0.25cm}, draw = none, at = {(0.5, 1.05)}, anchor = south }}
    ]
        \addplot[okabe_ito_1,ybar interval, ybar legend, no marks, fill, fill opacity={0.1}] table[col sep=comma, x=xhist, y=yhist] {\histogramfile};

        \addplot[okabe_ito_1,thick,no marks] table[col sep=comma, x=xpdf, y=gg_fit] {\histogramfile}
            node[pos=0.63, pin={[pin distance=7mm]45:{\(\sigma_I^2 \approx \qty{0.12}{}\)}}] {};

        \addplot[okabe_ito_3,ybar interval, ybar legend, no marks, fill, fill opacity={0.1}] table[col sep=comma, x=xhist, y=yhist] {\experimentfile};

        \addplot[okabe_ito_3,thick,no marks] table[col sep=comma, x=xpdf, y=gg_fit] {\experimentfile}
            node[pos=0.8, pin={[pin distance=7mm]45:{\(\sigma_I^2 \approx \qty{0.16}{}\)}}] {};

        \addlegendentry {Sim.}; 
        \addlegendentry {Sim. (Fit)}; 
        \addlegendentry {Exp.}; 
        \addlegendentry {Exp. (Fit)};

        \node[draw, fill=white, align=center, anchor=north east] at (rel axis cs:1,1) {Free-Space Powermeter};
    \end{axis}
\end{tikzpicture}%
\begin{tikzpicture}
    \def\histogramfile{figures/histograms/01_Cn2_1.5e-15_waveprop_power_in_fiber_HG00_modeLP01.csv}
    \def\experimentfile{figures/histograms/2026-01-27T03-50-00_fiber_power_histogram.csv}
    \tikzset{font=\footnotesize}
    \begin{axis}[
        width=0.5*\linewidth, 
        height=0.6*\axisdefaultheight, 
        axis lines={left}, 
        xlabel={{Normalized Power $P / \langle P\rangle$}}, 
        ylabel={{Probability density}}, 
        enlarge x limits={{rel=0.1, upper}}, 
        enlarge y limits={{rel=0.1, upper}}, 
        xmin=0, 
        xmax=3,
        ymin={0}, 
        grid, 
        legend columns={-1}, 
        legend style={{ /tikz/every even column/.append style={column sep=0.25cm}, draw = none, at = {(0.5, 1.05)}, anchor = south }}
    ]
        \addplot[okabe_ito_1,ybar interval, ybar legend, no marks, fill, fill opacity={0.1}] table[col sep=comma, x=xhist, y=yhist] {\histogramfile};

        \addplot[okabe_ito_1,thick,no marks] table[col sep=comma, x=xpdf, y=gg_fit] {\histogramfile}
            node[pos=0.53, pin={[pin distance=7mm]45:{\(\sigma_I^2 \approx \qty{0.14}{}\)}}] {};

        \addplot[okabe_ito_3,ybar interval, ybar legend, no marks, fill, fill opacity={0.1}] table[col sep=comma, x=xhist, y=yhist] {\experimentfile};

        \addplot[okabe_ito_3,thick,no marks] table[col sep=comma, x=xpdf, y=gg_fit] {\experimentfile}
            node[pos=0.55, pin={[pin distance=7mm]45:{\(\sigma_I^2 \approx \qty{0.178}{}\)}}] {};

        \addlegendentry {Sim.}; 
        \addlegendentry {Sim. (Fit)}; 
        \addlegendentry {Exp.}; 
        \addlegendentry {Exp. (Fit)};

        \node[draw, fill=white, align=center, anchor=north east] at (rel axis cs:1, 1) {Fiber Powermeter};
    \end{axis}
\end{tikzpicture}%
        \caption{Experimental (single-shot) versus simulated normalized power fading histograms for the 1-inch free-space detector and the SMF coupling at the median turbulence regime, i.\,e., at $C_{n}^{2}\approx \qty{1.5e-14}{m^{-2/3}}$. The histograms and estimated}
        \label{fig:histograms}
    \end{figure*}

    As shown in \autoref{fig:histograms}, the numerical statistics agree with the experimental data at the targeted median $C_{n}^{2}$ condition. To formalize this comparison, we fit a Gamma-Gamma distribution to the normalized fading histograms \cite{Andrews2005LaserEdition}. \autoref{tab:statistics} compares the extracted (power) scintillation index, \(\sigma_I^2=\frac{\langle P^2\rangle}{\langle P\rangle^2} - 1\), for both branches, where experimental values are determined from power measurements over the full 10-minute observation window corresponding to the scintillometer $C_n^2$ state. 
    The experimental free-space $\sigma_I^2$ slightly exceeds the simulated value, a variance likely attributable to short-term fluctuations in turbulence strength occurring within the 10-minute averaging window. The empirical fading in the SMF branch is more severely impacted by these fluctuations and is further degraded by the pointing-and-tracking system’s residual jitter. As our split-step model applies idealized tip-tilt correction, it intentionally isolates the fundamental atmospheric phase distortions from these hardware-specific mechanical limits. 

    Consequently, the digital twin establishes an optimistic theoretical baseline for turbulence-induced fading. Bounding both free-space amplitude scintillation and SMF coupling dynamics, this dual-branch validation establishes a baseline to quantify the spatial diversity gains of mode-diverse architectures.

    \begin{table}[]
        \centering
        \small
        \begin{tblr}
            {colspec={l|X[c]X[c]|X[c]X[c]},width=\linewidth} \toprule
            \SetCell[r=2]{m}$C_{n}^{2}/ \qty{}{\m^{-2/3}}$ & \SetCell[c=2]{c}Free-space PM & 2-3 & \SetCell[c=2]{c}Fiber PM & 2-5 \\
            & Sim. & Exp. & Sim. & Exp. \\
            \midrule \qty{1.5e-15}{} & \qty{0.12}{} & \qty{0.157}{} & \qty{0.14}{} & \qty{0.194}{} \\
            \qty{7.3e-15}{} & \qty{0.54}{} & \qty{0.78}{} & \qty{0.77}{} & \qty{1.15}{} \\
            \qty{1.5e-14}{} & \qty{1.03}{} & \qty{1.22}{} & \qty{1.83}{} & \qty{2.28}{} \\
            \bottomrule
        \end{tblr}
        \caption{Scintillation index $\sigma_{I}^{2}$ obtained from fitting a Gamma-Gamma distribution to the simulated and experimentally obtained fading statistics for both the free-space detector and the SMF coupling across the three targeted $C_{n}^{2}$ regimes. For the experimental data, the median across all available observations was considered.}
        \label{tab:statistics}
    \end{table}

    \section{Evaluating Mode-Diverse Reception}
    Using the validated complex amplitude distributions generated by our digital twin, we evaluate the achievable performance of mode-diverse receiver architectures employing $N$ spatial modes. In our emulation, the SMF of the Aircision terminal is replaced by few-mode step-index fibers supporting $N \in \{3, 6\}$ linearly polarized (LP) modes.

    The received field is demultiplexed into parallel single-mode coherent receivers using a lossless mode demultiplexer. Recent advances in mode multiplexers enable high efficiencies, supporting this approximation~\cite{PhotonicLanternsApplication2024,PROTEUSInventingOpticala}. 
    Assuming a receiver digital signal processing architecture capable of adaptive maximum-ratio combining (MRC), the weighted coherent summation of these individually coupled mode fields determines the effective signal quality.

    To evaluate reliability gains due to mode-diversity, we isolate the fiber coupling - the dominant dynamic impairment - while neglecting mitigable, static geometric collection losses. For each Monte Carlo iteration, we project the perturbed complex amplitude distributions onto the backpropagated fiber modes to extract the per-mode coupling efficiency $\eta_{i}$, yielding an effective diversity efficiency of $\eta_{\text{eff}}=\sum_i \eta_i$. 
    
    In order to achieve practicable reliability levels, the FSO system has to maintain a power margin beyond the receiver's sensitivity limit. After accounting for fixed hardware and path losses, the remaining power budget is allocated as a dynamic threshold $L_{\text{th}}$ defining the maximum weather- and turbulence-induced fading the link can endure. Applying this threshold to our Monte Carlo simulations of the effective coupling efficiency ($\eta_{\text{eff}}$) yields a discrete set of conditional outage probabilities for each mode count $N$:
    \begin{equation}
        P_\text{out}(C_{n}^{2}) = P(L > L_{\text{th}}| C_{n}^{2})
    \end{equation}

\begin{figure}[b]
        \centering
\begin{tikzpicture}
    \def\datafile{figures/outage/Pout_vs_cn2.csv}
    \def\cn2file{figures/outage/cn2_kde.csv}
    \tikzset{font=\footnotesize}
    \begin{axis}[
        width=\linewidth, 
        height=0.7*\axisdefaultheight, 
        axis lines={left}, 
        xlabel={{Refractive Index Structure Constant $C_n^2/\qty{}{m^{-2/3}}$}}, 
        ylabel={{Outage Probability $P_\text{out}(L_\text{th}\,|\,C_n^2)$}}, 
        enlarge x limits={{rel=0.1, upper}}, 
        enlarge y limits={{rel=0.1, upper}}, 
        xmode=log,
        ymode=log,
        xmin={1.5e-16},
        xmax={1.5e-13},
        ymin={1e-3}, 
        grid, 
        legend columns={-1}, 
        legend style={{ /tikz/every even column/.append style={column sep=0.25cm}, draw = none, at = {(0.5, 1.05)}, anchor = south }}
    ]
        \addplot+[thick, mark=*] table[col sep=comma, x=x, y=y1] {\datafile};
        \addplot+[thick,mark=x] table[col sep=comma, x=x, y=y3] {\datafile};
        \addplot+[thick,mark=o] table[col sep=comma, x=x, y=y6] {\datafile};

        \addlegendentry {SMF}; 
        \addlegendentry {FMF3}; 
        \addlegendentry {FMF6}; 

        \node[draw, fill=white, align=center, anchor=north west] at (rel axis cs:0.025, 1) {\(L_\text{th}=\qty{10}{dB}\)};
    \end{axis}
\end{tikzpicture}%
        \caption{Simulated conditional outage probability $P_\text{out}(C_{n}^{2})$ versus $C_{n}^{2}$ for SMF, 3-mode, and 6-mode fiber receivers.}
        \label{fig:outage_conditional}
        \label{fig:outage}
    \end{figure}
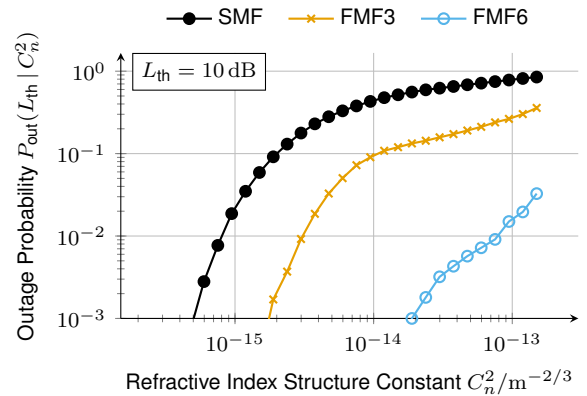

    In this context, we run the Monte Carlo simulations for a wide range of $C_{n}^{2}$ values, spanning from \qty{1.5e-16}{\m^{-2/3}} to \qty{1.5e-13}{\m^{-2/3}}, to capture the full spectrum of outage-relevant turbulence conditions observed in the long-term scintillometer data. From these discrete operational points, we obtain an estimate on the conditional outage probability for each mode assuming a loss threshold \(L_\text{th}=\qty{10}{\dB}\), as shown in \autoref{fig:outage}. 
    Numerically integrating over the product of the conditional outage probabilities and the probability density function of $C_{n}^{2}$ obtained from the scintillometer data, we determine the long-term outage probability for each mode count $N$ across a wide range of turbulence strengths:
    \begin{equation}
        P_{\text{out}}= \int P_\text{out}(C_{n}^{2}) \cdot p(C_{n}^{2}) \, \text{d}C_{n}^{2}
    \end{equation}
    
    Whereas for single-mode operation, the total outage probability becomes \qty{0.131}{}, incorporating three and six modes reduces this long-term outage floor to \qty{1.76e-2}{} and \qty{2.02e-5}{}, respectively.

    \section{Conclusion}
    We established and validated a numerical model for turbulence-induced fading statistics targeting a representative 4.6-km-long FSO link. With power measurements isolating complex amplitude and pure-amplitude perturbations we have verified that our numerical split-step channel model describes both large-aperture irradiance and fiber coupling fading.
    Based on long-term scintillometer data, we have then demonstrated the potential of mode-diverse receivers to significantly reduce the long-term outage probability due to turbulence-induced fading by four orders of magnitude.
    \clearpage
    \section{Acknowledgements}
    Supported by the European Innovation Council Transition project CombTools (G.A. 101136978), the Dutch Research Council (NWO) TTW-Perspectief Optical Wireless Superhighways: Free photons (at home and in space): FREE P19-13, and the Dutch Ministry of Economic Affairs and Climate Policy (EZK) via the PhotonDelta National Growth Fund Programme on Photonics. 




    \printbibliography

    \vspace{-4mm}

\end{document}